\documentclass[aps,prb,superscriptaddress,twocolumn]{revtex4-2}
\usepackage[normalem]{ulem}
\usepackage[cp1251]{inputenc}
\usepackage{amsmath}
\usepackage{amssymb}
\usepackage{amsfonts}
\usepackage{color}
\usepackage{graphicx}
\usepackage{lipsum,graphicx}
\usepackage{amsfonts}
\usepackage{textcomp}
\usepackage{hyperref}
\usepackage{hyperref}

\newcommand{\beq}{\begin{equation}}
\newcommand{\eeq}{\end{equation}}
\newcommand{\bea}{\begin{eqnarray}}
\newcommand{\eea}{\end{eqnarray}}

\begin{document}

\title{
	Exchange Splitting Mechanism of  Negative Magnetoresistance in Layered Antiferromagnetic
	Semimetals}
\author{P. D. Grigoriev}
\affiliation{L.D. Landau Institute of Theoretical Physics, RAS}
\affiliation{National University of Science and Technology ''MISiS'', 119049, Moscow, Russia} 
\affiliation{HSE University, Moscow 101000, Russia}
\author{N. S. Pavlov}
\affiliation{Institute for Electrophysics,  RAS, Ekaterinburg, 620016, Russia}
\author{I. A. Nekrasov}
\affiliation{Institute for Electrophysics, RAS, Ekaterinburg, 620016, Russia}
\author{I. R. Shein}
\affiliation{Institute of Solid State Chemistry, RAS, Ekaterinburg, 620990, Russia}
\author{A. V. Sadakov}
\author{O. A. Sobolevskiy}
\affiliation{V.~L.~Ginzburg Research Center at P.~N.~Lebedev Physical Institute, RAS, Moscow 119991, Russia}
\author{E.  Maltsev}
\affiliation{Leibniz Institute for Solid State and Materials Research, IFW Dresden, 
	D-01069 Dresden, Germany}
\affiliation{Dresden-W\"{u}rzburg Cluster of Excellence ct.qmat, Dresden, Germany}
\author{N. P\'{e}rez}
\affiliation{Leibniz Institute for Solid State and Materials Research, IFW Dresden, 
D-01069 Dresden, Germany}
\author{L. Veyrat}
\affiliation{Leibniz Institute for Solid State and Materials Research, IFW Dresden, 
	D-01069 Dresden, Germany}
\affiliation{Dresden-W\"{u}rzburg Cluster of Excellence ct.qmat, Dresden, Germany}
\affiliation{Laboratoire National des Champs Magnetiques Intenses, CNRS-INSA-UJF-UPS,  F-31400 Toulouse, France}
\author{V. M. Pudalov}
\affiliation{V.~L.~Ginzburg Research Center at P.~N.~Lebedev Physical Institute, RAS,  Moscow 119991, Russia}

\begin{abstract}
Layered topologically non-trivial and trivial semimetals with AFM-type
ordering of magnetic sublattice are known to exhibit a negative
magnetoresistance that is well correlated with AFM magnetization changes in
a magnetic field. This effect is reported in several experimental studies
with 
EuFe$_2$As$_2$, EuSn$_2$As$_2$,  EuSn$_2$P$_2$, etc., where the resistance decreases quadratically with
field by about $\delta\rho/\rho \sim 4-6\%$ up to the spin-polarization field.
Despite the fact that this effect is well documented experimentally, its theoretical
explanation is missing up to date. In this paper we propose 
novel theoretical mechanism describing the observed magnetoresistance that
does not imply either topological origin of the materials, surface
roughness, their potential defect structure, or electron-magnon scattering.
We believe, the proposed intrinsic mechanism of magnetoresistance is
applicable to a wide class of the layered AFM- ordered semimetals. The
theoretically calculated magnetoresistance is qualitatively consistent with experimental
data for crystals  of various composition.
\end{abstract}

\maketitle

	Topology incorporated with magnetism provides a fertile playground for studying novel quantum states 
	and  therefore attracts tremendous attention. 	Topological aspects and related electronic phenomena  
	occuring in antiferromagnetic (AFM) materials are among the central topics in 
	condensed matter physics. 
	
	The vast majority of AFM topologically nontrivial insulators (TI) and  semimetals (TSM)  exhibit  such 
	characteristic feature as a negative magnetoresistance (NMR) in  a magnetic fields parallel to the basal plane.
	This effect has recently attracted  a great deal of interest,
	as it is 	related with chiral anomaly in Dirac  and Weyl semimetals 
	\cite{li_NatCom_2015, yan_PRR_2022}, with 
Chern insulators and  anomalous quantized Hall state in AFM topological insulators 
	\cite{li_PRB_2019, ge_NatlSciRev_2020}.
		Indeed, for topological semimetals (Na$_3$Bi, TaS, Cd$_3$As$_2$), a 
	large NMR results from the chiral anomaly  \cite{Xiong, HuangTaAsNMR2015} 
	and appears only when the electric field is applied nearly parallel to the magnetic field.
	This  NMR is highly anisotropic and its sign changes as the angle between $\mathbf{H}$ and current 
	increases to $\pi/2$ \cite{Xiong}. In the topological insulator (TI) MnBi$_2$Te$_4$, 
	a huge NMR \cite{ge_NatlSciRev_2020}  occurs as a result of the  
	topological phase transition  from  AFM TI to a ferromagnetic Weyl semimetal in external field.

		On the other hand, a large number of  topologically trivial analogues,  the layered AFM semimetals, 
	 also exhibit NMR. 	There is no widely accepted treatment of NMR. 
	Particularly,  in doped magnetic materials 
	it is  being associated with  scattering by magnon and impurities, whereas topology is evidently irrelevant.
	
	In this paper we suggest a novel mechanism of the {\em intrinsic} NMR  that is irrelevant to magnons, impurities,
	and topology 	and we believe is applicable to a  wide class of  non-Weyl semimetals, such as 
	 EuSn$_2$As$_2$, EuFe$_2$As$_2$, EuSn$_2$P$_2$. We test our model by 
	comparing it with experimentally measured magnetoresistance in  EuSn$_2$As$_2$. 

 This paper  focuses on the topologically  trivial layered 
semimetals with AFM-type ordering of the sublattice of magnetic ions. These materials
exhibit  a negative magnetoresistance that is tightly correlated with the magnetization field dependence.
The  magnetization changes linearly with external field in conventional manner   and 
 sharply  saturates above a  field of complete spin polarization  $H_{\rm sf}$. The combination of the two 
 closely related effects was reported in several experimental studies with 
EuFe$_2$As$_2$  \cite{jiang_NJP_2009, sanchez_PRB_2021}, EuSn$_2$As$_2$, and EuSn$_2$P$_2$ \cite{gui_ACSCentSci_2019}.

\begin{figure}[tbp]
\includegraphics[width=150pt]{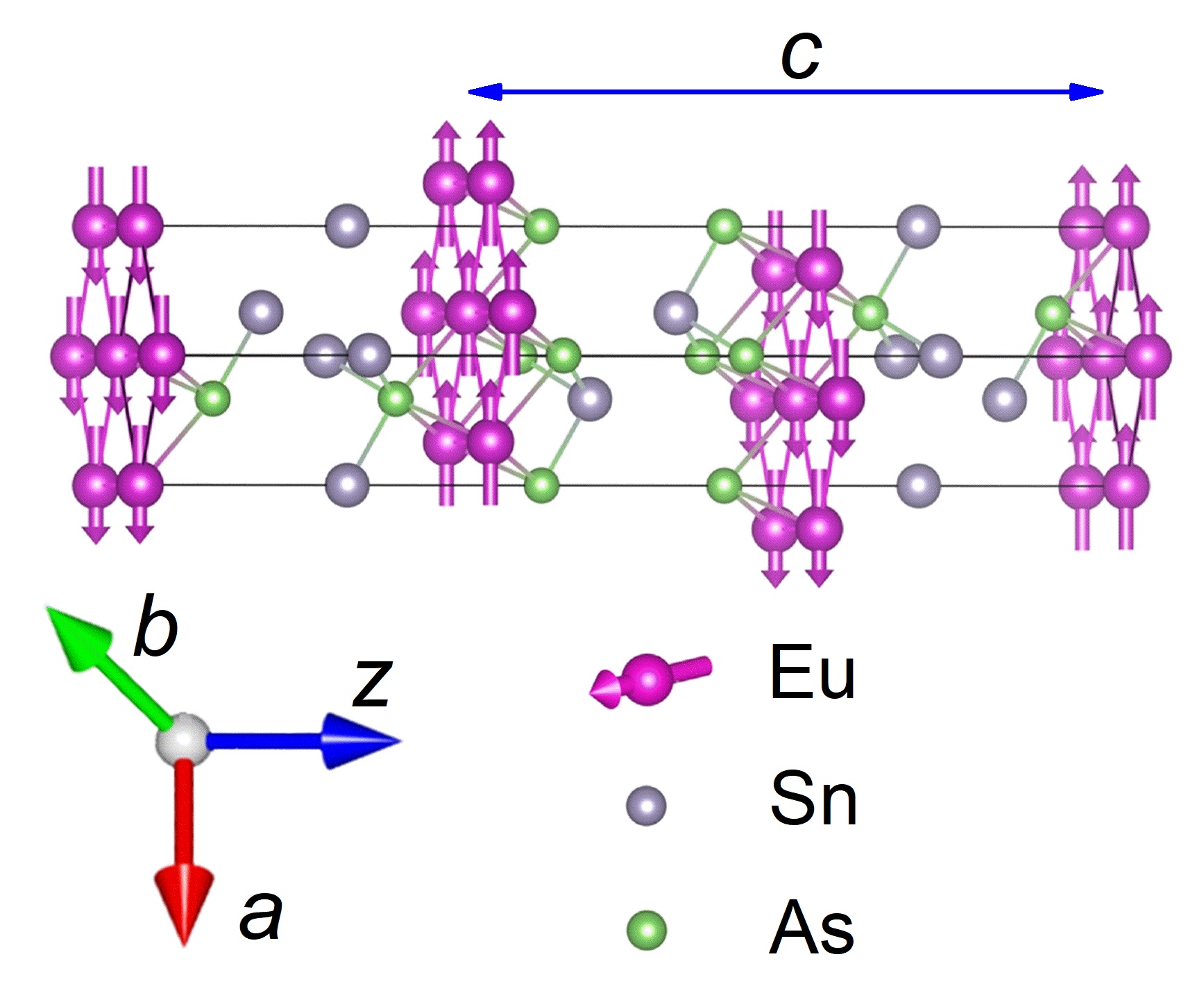}  
	\caption{Schematic picture of the EuSn$_2$As$_2$ lattice structure (adapted from Ref.~\cite{golov_PRB_2022}).
		Magenta arrows show Eu-atoms magnetization direction in the A-type AFM
		ordered state.  Horizontal arrow  denotes the lattice spacing $c\approx 2.64$nm along $z$-axis   \cite{arguilla_InChemFront_2017, chen_ChPhysLet_2020, PRB_tbp}. 
	}
	\label{fig:1}
\end{figure}

For the particular representative EuSn$_2$As$_2$ compound, 
at $T<24$K the magnetic Eu-sublattice experiences magnetic  ordering into the
A-type AFM structure in which Eu magnetic moments lie in the easy $ab$-plane and
rotate by $\pi$ from layer to layer along $z$-axis (see Fig.~\ref{fig:1}a). 

Synchronous with  magnetization, the diagonal resistivity 
decreases approximately parabolically with field as $\delta\rho_{xx}(H) \propto -\alpha H^2$. 
At higher fields $H>H_{\rm sf}$,  magnetotransport perpendicular to the field sharply changes from NMR for a conventional positive magnetoresistance.
 Such NMR closely correlated 
with the magnetization $M(H)$ field dependence \cite{jiang_NJP_2009, chen_ChPhysLet_2020, li_PRB_2021, gui_ACSCentSci_2019} 
was discussed previously in terms of either nontrivial topological properties, or in terms
of scattering by domains, grain boundaries etc., with no detailed theoretical consideration.

 In this paper we address the issue of the origin of negative parabolic MR. Specifically,  
 	(i) we propose a model where NMR universally originates from the enhancement of electron scattering in AFM crystals,   
	(ii) we substantiate this proposal using DFT calculations of the spin polarized charge distribution, 
		and (iii) we compare the model with our experimental data for NMR.

\begin{figure}[tbh]
\includegraphics[width=80pt]{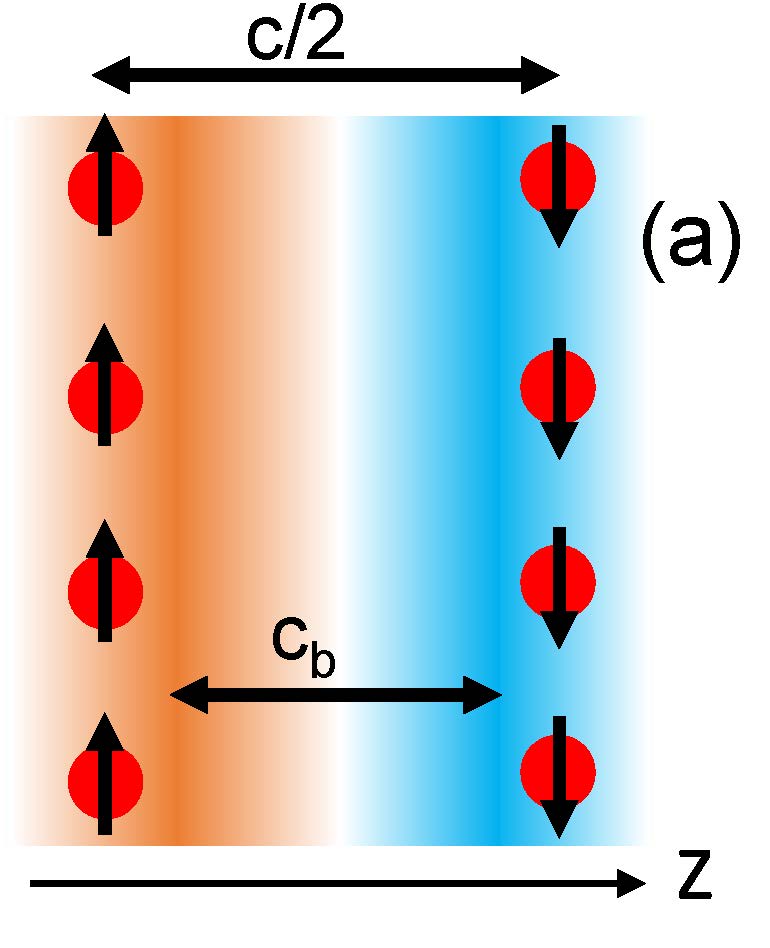} 
\includegraphics[width=140pt]{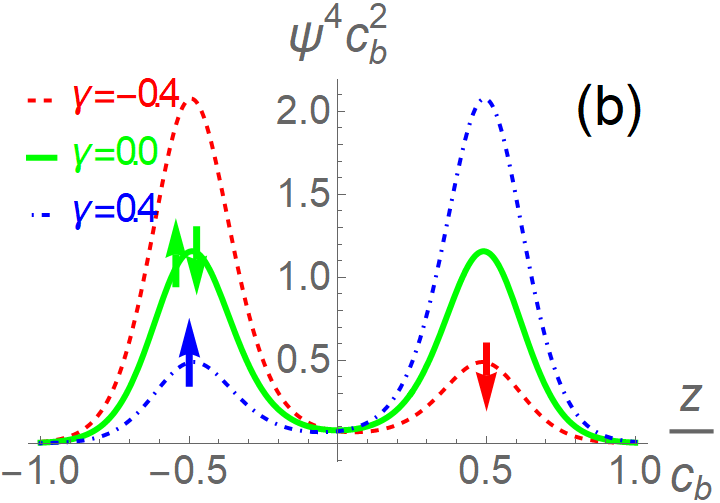}
\caption{Wave functions magnitude distribution along the crystal $z$-axis, 
	for the lowest-energy quantum state [Eq.~(\protect\ref{psipmWF})] in a
	double-well potential, modeling two AFM\ sublattices. 
	(a) Schematic color illustration of the spin-up (red) and spin-down (blue) 
wave function distribution along the $z$-axis in the half-cell of the $c/2$-size. 
$c_b$ denotes the distance between the two spin-split WF maxima.   
Color intensity represents
the WF magnitude. Circles show Eu-atoms, and arrows - their magnetization direction in the AFM state. 
(b)  Schematic picture of the fourth power $%
\protect\psi ^{4}(z)$, entering the scattering rate in Eq. (\protect\ref%
{tau}). The asymmetry parameter is $\protect\gamma \equiv \Delta E_{Zex}/2t_0 =0$
(green solid line), $\protect\gamma=-0.4$ (dashed red line) and $\protect%
\gamma=0.4$ (dash-dotted blue line). 
}
\label{FigWF}
\end{figure}

The main idea of the proposed NMR mechanism is as follows. 
The AFM order violates the $\hat{Z}_{2}$ symmetry $1\leftrightarrow 2$ 
between two magnetic sublattices (see Fig. 1a). This squeezes
the electron wave function $\psi _{\sigma }\left( \boldsymbol{r}\right) $
for each spin component $\sigma $ and, hence, increases the electron
scattering rate \cite{Abrik} 
\begin{equation}
1/\tau \propto \int \left\vert \psi _{\sigma }\left( \boldsymbol{r}\right)
\right\vert ^{4}d^{3}\boldsymbol{r}  \label{tau}
\end{equation}%
by short-range crystal defects or $\delta$-correlated disorder already in the Born approximation.
 Figures \ref{FigWF}a,b schematically illustrate the violation of $\hat{Z}_{2}$ symmetry of
the electron wave function by AFM order and the enhancement of $\left\vert
\psi \right\vert ^{4}(\boldsymbol{r})$ entering the scattering rate in Eq. (%
\ref{tau}). An external magnetic field $\boldsymbol{H}$ below the  complete spin-polarization  
and  perpendicular to the AFM sublattice magnetization $M_{AFM}$ reduces it approximately according to classical relation \cite{kittel}
\begin{equation}
\frac{M_{AFM}\left( H\right) }{M_{AFM}\left( 0\right) }\approx \sqrt{1-\frac{%
M_{H}^{2}}{M_{AFM}^{2}\left( 0\right) }}\approx \sqrt{1-\frac{H^{2}}{H_{%
\mathrm{sf}}^{2}}},  \label{M}
\end{equation}%
where $M_{H}=\chi H$ is the field-induced magnetization along the magnetic
field and perpendicular to AFM order.  We took $\chi =\chi _{\perp }(H)\approx const$, in agreement
with textbook \cite{kittel} and with experimental data for EuSn$_2$As$_2$ 
\cite{arguilla_InChemFront_2017, chen_ChPhysLet_2020, pakhira-EuMg2Bi2_PRB_2021}.
As we show below, the degree of $\hat{Z}_{2}$ symmetry
violation and $1/\tau $ enhancement depends on the ratio 
\begin{equation}
\gamma =\Delta E_{ex}/2t_{0}  \label{delta}
\end{equation}%
of the exchange splitting $\Delta E_{ex}$ of  conduction electron bands to their
hopping amplitude $t_{0}$ between the opposite AFM sublattices  (Fig.~\ref{fig:1}a). 
While $t_{0}$ is
determined by the band structure and does not considerably depend on $H$,
the exchange splitting $\Delta E_{ex}\propto M_{AFM}\left( H\right) $
decreases with $H$ according to Eq. (\ref{M}): 
\begin{equation}
\frac{\Delta E_{ex}(H)}{\Delta E_{ex}(H=0)}=\frac{M_{AFM}\left( H\right) }{%
M_{AFM}\left( 0\right) }\approx \sqrt{1-\left( H/H_{\mathrm{sf}}\right) ^{2}}%
.  \label{EexB}
\end{equation}%
This leads to a 
 parabolic negative magnetoresistance almost up to the 
 field $H_{\mathrm{sf}}$.   Closer to the 
	spin polarization transition, at $H\rightarrow H_{\mathrm{sf}}$ a
first-order spin-flip phase transition  causes a much faster decrease of $%
M_{AFM}\left( H\right) $ than that given by Eq. (\ref{M}).

Let us consider the electron wave functions squeezing in the AFM state.  The two AFM
sublattices are numerated by the index $i=1,2$ or by the \textquotedblright
pseudospin\textquotedblright\ $\lambda =\lambda \left( i\right) =3/2-i$,
while the projection of spin on the magnetization axis $\boldsymbol{M}_{AFM}$
is denoted by $\sigma =\pm 1\equiv \uparrow ,\downarrow $. The quantum basis
consists of four states, $\left\vert i,\sigma \right\rangle  = \left\{ 1\uparrow ,1\downarrow ,2\uparrow
,2\downarrow \right\}$, corresponding to wave functions $\psi _{i,\sigma
}=\left\{\psi _{1\uparrow },\psi _{1\downarrow },\psi _{2\uparrow },\psi
_{2\downarrow }\right\} $. The usual Zeeman splitting $\Delta E_{Z}(%
\boldsymbol{H})=\left( \boldsymbol{\vec{\sigma}\cdot \vec{H}}\right) g\mu
_{H}/2$ in a relevant external magnetic field $H\lesssim 5$T is much smaller
than the exchange splitting $\Delta E_{ex}\gtrsim 10$meV and will be
neglected below. Then the AFM-sublattice part of the electron Hamiltonian
for each electron quasi-momentum $\boldsymbol{k}$ is given by the $4\times 4$
matrix, which decouples into two $2\times 2$ matrices: 
\begin{equation}
\hat{H}_{\sigma }=\left( 
\begin{array}{cc}
\Delta E_{ex}\sigma /2 & t_{0} \\ 
t_{0}^{\ast } & -\Delta E_{ex}\sigma /2%
\end{array}%
\right) .  \label{HBLs}
\end{equation}%
Here the non-diagonal term $t_{0}=t_{0}^{\ast }$\ is the intersublattice
electron transfer integral. The diagonalization of Hamiltonian (\ref{HBLs})
gives two eigenvalues 
\begin{equation}
E_{\pm ,\sigma }=\mp \sqrt{\Delta E_{ex}^{2}/4+t_{0}^{2}}  \label{Epm}
\end{equation}%
and the corresponding wave functions (WF)
\begin{equation}
\psi _{\pm ,\sigma }=\frac{\psi _{1}\left( \sigma \gamma\pm \sqrt{\gamma
^{2}+1}\right) +\psi _{2}}{\sqrt{1+\left( \sigma \gamma\pm \sqrt{\gamma
^{2}+1}\right) ^{2}}},  \label{psipmWF}
\end{equation}%
where $\psi _{1},_{2}$ are the electron wave functions ``localized'' mostly on
the first and second AFM sublattices. 
Without the AFM order, i.e. at $\Delta E_{ex}=0$, one gets the electron
spectrum $E_{\pm ,\sigma }=\mp t_{0}$ and the corresponding normalized
eigenstates 
\begin{equation}
\psi _{\pm ,\sigma }^{s}=\left( \psi _{1}\pm \psi _{2}\right) /\sqrt{2},
\label{psipmS}
\end{equation}%
which are symmetric and antisymmetric superpositions of electron states on
sublattices $1$ and $2$, as it should be when the $\hat{Z}_{2}$ symmetry is
conserved. From Eqs. (\ref{psipmWF}) we see that AFM lifts this symmetry,
 making the eigenfunction  amplitude larger on one of two
sublattices, as illustrated in Fig. \ref{FigWF}. This enhances the integral in 
Eq. (\ref{tau}), as seen from  Figs. \ref{FigWF} and shown below. 

In the Born approximation, i.e. the second-order perturbation theory in
the impurity potential, the electron scattering rate is given by the Fermi's
golden rule \cite{Abrik}:%
\begin{equation}
\frac{1}{\tau }=\frac{2\pi }{\hbar }\sum_{n^{\prime },i}\left\vert T_{ 
\boldsymbol{n}^{\prime }\boldsymbol{n}}^{(i)}\right\vert ^{2}\delta\left(
\varepsilon _{\boldsymbol{n}}-\varepsilon _{\boldsymbol{n}^{\prime }}\right)
.  \label{TauGFR}
\end{equation}%
where the index $n\equiv \left\{ \boldsymbol{k},\zeta ,\sigma \right\} $
numerates quantum states, $\zeta $ and $\sigma $
denote the electron subband and spin projection,  
$\varepsilon _{\boldsymbol{n}}$ is the electron energy in state $n$, and $\delta (x)$ is the Dirac delta-function. 
The short-range impurities or other crystal defects in
solids are, usually, approximated by the point-like potential 
$V_{i}\left( \boldsymbol{r}\right) =U\delta\left( \boldsymbol{r}-%
\boldsymbol{r}_{i}\right) $.
Here we omit the spin index $\sigma $ because it is conserved by
the potential scattering.  The corresponding matrix
element of electron scattering by this impurity potential is
$T_{\boldsymbol{n}^{\prime }\boldsymbol{n}}^{(i)}=U 
\Psi^{\ast } _{\boldsymbol{n}^{\prime }}\left( \boldsymbol{r}_{i}
\right) \Psi _{\boldsymbol{n}}\left( \boldsymbol{r}_{i}
\right) $, where $\Psi _{\boldsymbol{n}}\left( \boldsymbol{r}\right) $ is 
the electron wave function in the state $\boldsymbol{n}$.

For short-range impurities the matrix element does not depend on electron
momentum $\boldsymbol{k}^{\prime }$. Then the summation over $\boldsymbol{k}%
^{\prime }$ with the $\delta $-function in Eq. (\ref{TauGFR}) gives the
electron density of states (DoS) $\nu _{F\zeta }\equiv \nu _{\zeta }\left(
\varepsilon _{F}\right) $ at energy Fermi $\varepsilon _{F}$ per one
spin component and per one subband $\zeta $. The total DoS per one spin
$\nu _{F}=\sum_{\zeta }\nu _{F\zeta }$. If the impurities are
uniformly and randomly distributed in space, the sum over impurities
rewrites as an integral over impurity coordinate: $\sum_{i}\rightarrow
n_{i}\int d^{3}\boldsymbol{r}_{i}$, where $n_{i}$ is the impurity
concentration. Then Eq. (\ref{TauGFR}) becomes 
\begin{equation}
\frac{1}{\tau }=\frac{2\pi }{\hbar }n_{i}U^{2}\int d^{3}\boldsymbol{r}%
_{i}\sum_{\zeta ^{\prime }}\nu _{F{\zeta }^{\prime }}\left\vert \Psi _{%
\boldsymbol{k}^{\prime }\zeta ^{\prime }}^{\ast }\left( \boldsymbol{r}%
_{i}\right) \Psi _{\boldsymbol{k}\zeta }\left( \boldsymbol{r}_{i}\right)
\right\vert ^{2}.  \label{Tau1}
\end{equation}%
The Bloch wave function is $\Psi _{\boldsymbol{k}\zeta }\left( \boldsymbol{r}%
\right) =\psi _{\zeta }\left( \boldsymbol{r}\right) \exp \left( i\boldsymbol{%
kr}\right) $, where $\psi _{\zeta }\left( \boldsymbol{r}\right) $ is the
periodic function. Then Eq. (\ref{Tau1}) rewrites as%
\begin{equation}
\frac{1}{\tau }=\frac{2\pi }{\hbar }n_{i}U^{2}\nu _{F}\,I,  \label{Tau2}
\end{equation}%
where the integral over one elementary cell
\begin{equation}
I\equiv \int d^{3}\boldsymbol{r}\left\vert \psi _{\zeta }\left( \boldsymbol{r%
}\right) \right\vert ^{2}\sum_{\zeta ^{\prime }}\frac{\nu _{F\zeta ^{\prime
}}}{\nu _{F}}\left\vert \psi _{\zeta ^{\prime }}\left( \boldsymbol{r}\right)
\right\vert ^{2} . \label{I}
\end{equation}%
For  a single band $\zeta $ at the Fermi
level, i.e. when $\nu _{F\zeta ^{\prime }}=0$ for $\zeta ^{\prime }\neq
\zeta $, this confirms Eq. (\ref{tau}). Eqs. (\ref{Tau2}) and (\ref{I}) also
result to Eq. (\ref{tau}) if only the scattering within the same band is
allowed, e.g.,  due to the spin conservation during the potential scattering. 
Eqs. (\ref{Tau2}),(\ref{I}) approximately
give Eq. (\ref{tau}) also when there are several bands, but the DoS $\nu _{F%
\boldsymbol{\zeta }}$ for one band $\zeta $ is much larger than that for the
others. The latter happens in Dirac semimetals, where the
DoS $\nu _{\zeta 
}\left( \varepsilon _{F}\right) \propto \varepsilon _{F}^{d-1}$ strongly
depends on the Fermi energy $\varepsilon _{F}$ and where the energy difference $%
\varepsilon _{\zeta ^{\prime }}\left( \boldsymbol{k}\right) -\varepsilon
_{\zeta }\left( \boldsymbol{k}\right) \gtrsim \varepsilon _{F}$ for $\zeta
^{\prime }\neq \zeta $.

Now we estimate the difference $\delta I$ of two integrals (\ref{I}) for the
asymmetric and symmetric wave functions, given by Eqs. (\ref{psipmWF}) and (\ref{psipmS}). 
The host crystal lattice has the $\hat{Z}_2$ symmetry,  
therefore $\int \psi _{1}^{4}\left( z\right) dz=\int \psi _{2}^{4}\left( z\right) dz$.
For simplicity, we also assume that the overlap of the wave functions on
different AFM sublattices is negligible, i.e. $\psi _{1}\psi _{2}\ll
\left\vert \psi _{1}\right\vert ^{2}$, and we neglect the products $\psi
_{1}\psi _{2}\approx 0$. The main conclusion remains valid also when $\psi
_{1}\psi _{2}\sim \left\vert \psi _{1}\right\vert ^{2}$, but the calculation
are more cumbersome.

First we consider the completely subband-polarized case $\nu _{F\boldsymbol{%
\zeta }}/\nu _{F}=1$ for $\zeta =1$ and $\nu _{F\boldsymbol{\zeta }}/\nu
_{F}=0$ for $\zeta =2$. Then for the symmetric wave function (\ref{psipmS}),
corresponding to the lowest subband $\zeta $ without the AFM order and
bilayer asymmetry, the integral (\ref{I}) is 
\begin{equation}
I_0=\int d^{3}\boldsymbol{r\,}\left\vert \psi
_{+}^{s}\right\vert ^{4}dz\approx \int d^{3}\boldsymbol{r\,}\psi
_{1}^{4}/2.  \label{I0}
\end{equation}%
We now calculate the difference 
\begin{equation}
\delta I\equiv I-I_0=\int d^{3}\boldsymbol{r\,}\left( \left\vert
\psi _{+}\right\vert ^{4}-\left\vert \psi _{+}^{s}\right\vert
^{4}\right) ,  \label{dI0}
\end{equation}%
which gives the correction to mean free time $\tau $ according to Eq. (\ref%
{tau}) or (\ref{Tau2}). After the substitution of Eqs. (\ref{psipmWF}) and (\ref{psipmS}), 
at $\psi _{1}\psi _{2}\ll \left\vert \psi _{1}\right\vert
^{2}$ this simplifies to  
\begin{equation}
\delta I \approx \boldsymbol{\,}\frac{\gamma ^{2}}{ 1+\gamma ^{2}}
\int d^{3}\boldsymbol{r}\frac{\psi _{1}^{4}\left( z\right) }{2}=
\frac{\gamma ^{2}\,I_0}{ 1+\gamma ^{2}} =
\frac{\gamma ^{2}\,I}{ 1+2\gamma ^{2}}.  \label{dI}
\end{equation}%

At $\gamma ^{2}\ll 1$ Eq. (\ref{dI}) simplifies to $\delta I\approx \gamma ^{2}I_0\approx \gamma ^{2}I$,
and substituting  Eqs. (\ref{dI}), (\ref{delta}) and (\ref{EexB}) to (\ref{Tau2}) we obtain the relative
increase of resistivity due to the AFM ordering%
\begin{equation}
\frac{\delta \rho (H)}{\rho (0)}
\approx \frac{\delta I}{I}\approx \frac{[\delta(H)]^{2}}{1+2[\delta(H)]^{2}} \approx \left( \frac{\Delta
E_{ex}}{2t_{0}}\right)^{2}\left( 1-\frac{H^{2}}{H_{\mathrm{sf}}^{2}}%
\right).  \label{drho}
\end{equation}%
Contrary to the NMR caused by  
chiral anomaly in Weyl semimetals, this increase of
resistivity is {\em isotropic}. For example, our NMR mechanism applies both
for the interlayer and in-plane current directions in layered conductors,
and it only slightly depends on the magnetic field direction due  to a 
magnetic anisotropy solely.

\begin{figure}[tbh]
\includegraphics[width=111pt]{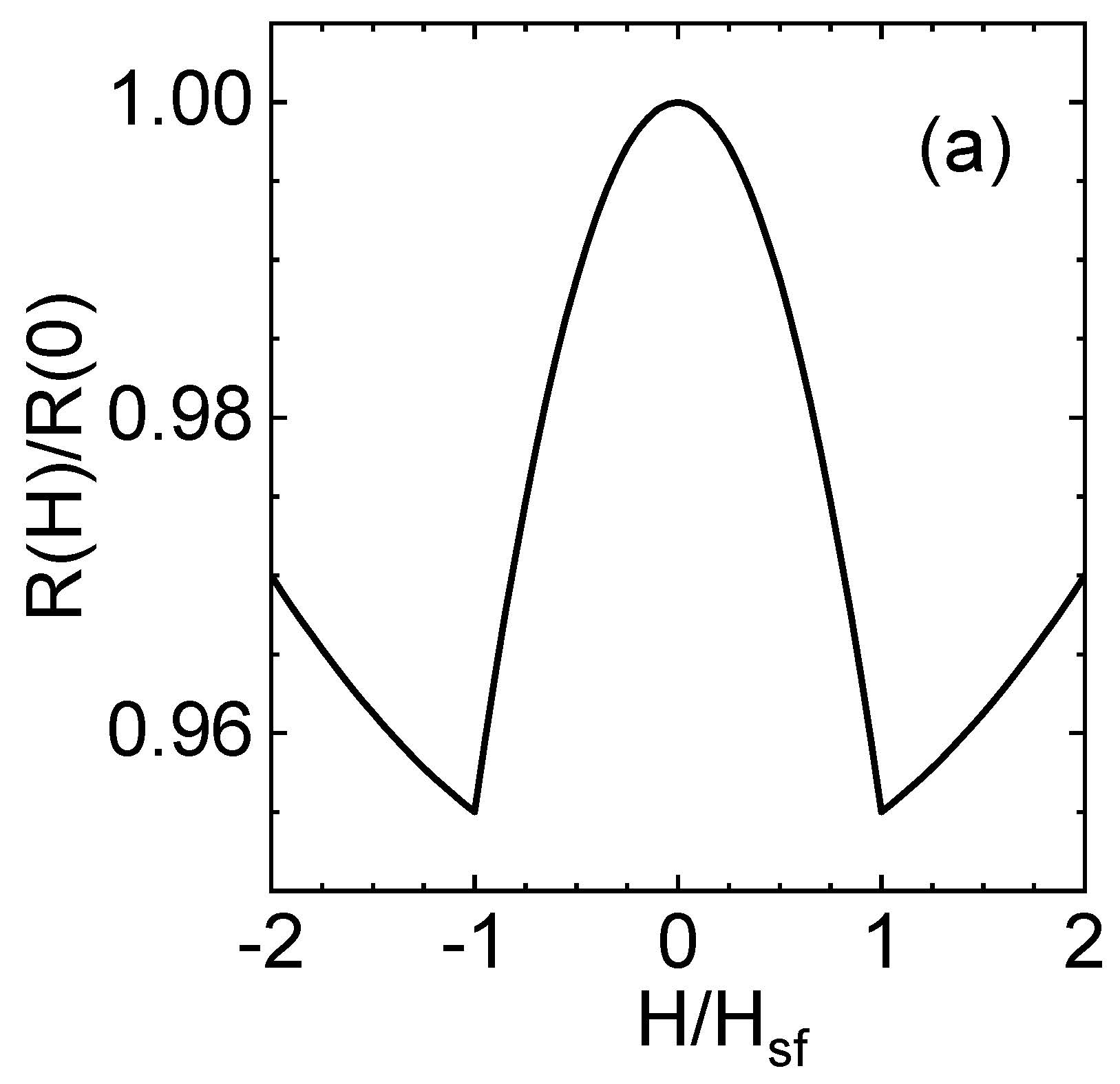}
\includegraphics[width=110pt]{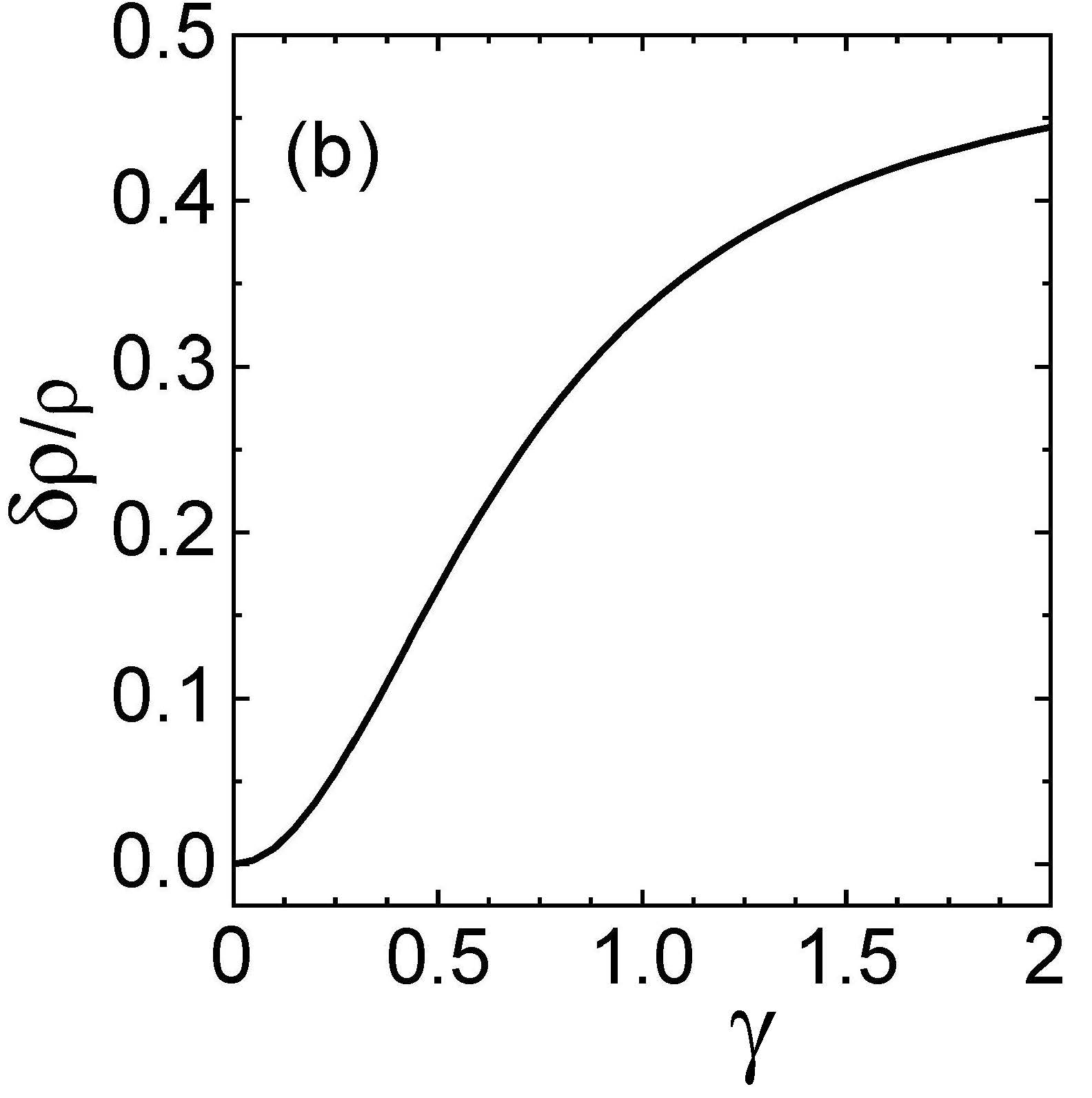}  
\caption{(a) Magnetoresistance curve given by Eq. (\protect\ref{RMzF}) at $%
\Delta E_{Zex}/2t_z =0.1$ and $[\protect\omega_c\protect\tau 
(H=H_{\mathrm{sf}})]^2 =0.2$. (b) The maximal possible relative value 
of the proposed NMR effect as a  function of $\gamma=\Delta E_{ex}/2t_{0} $, plotted using Eqs.  (\ref{dI}) or (\ref{drho}). 
It saturates to $50$\%  at $\gamma\gg 1$, when resistivity drops by half.
}
\label{FigRB}  
\end{figure}

\begin{figure}[ht]
\includegraphics[width=120pt]{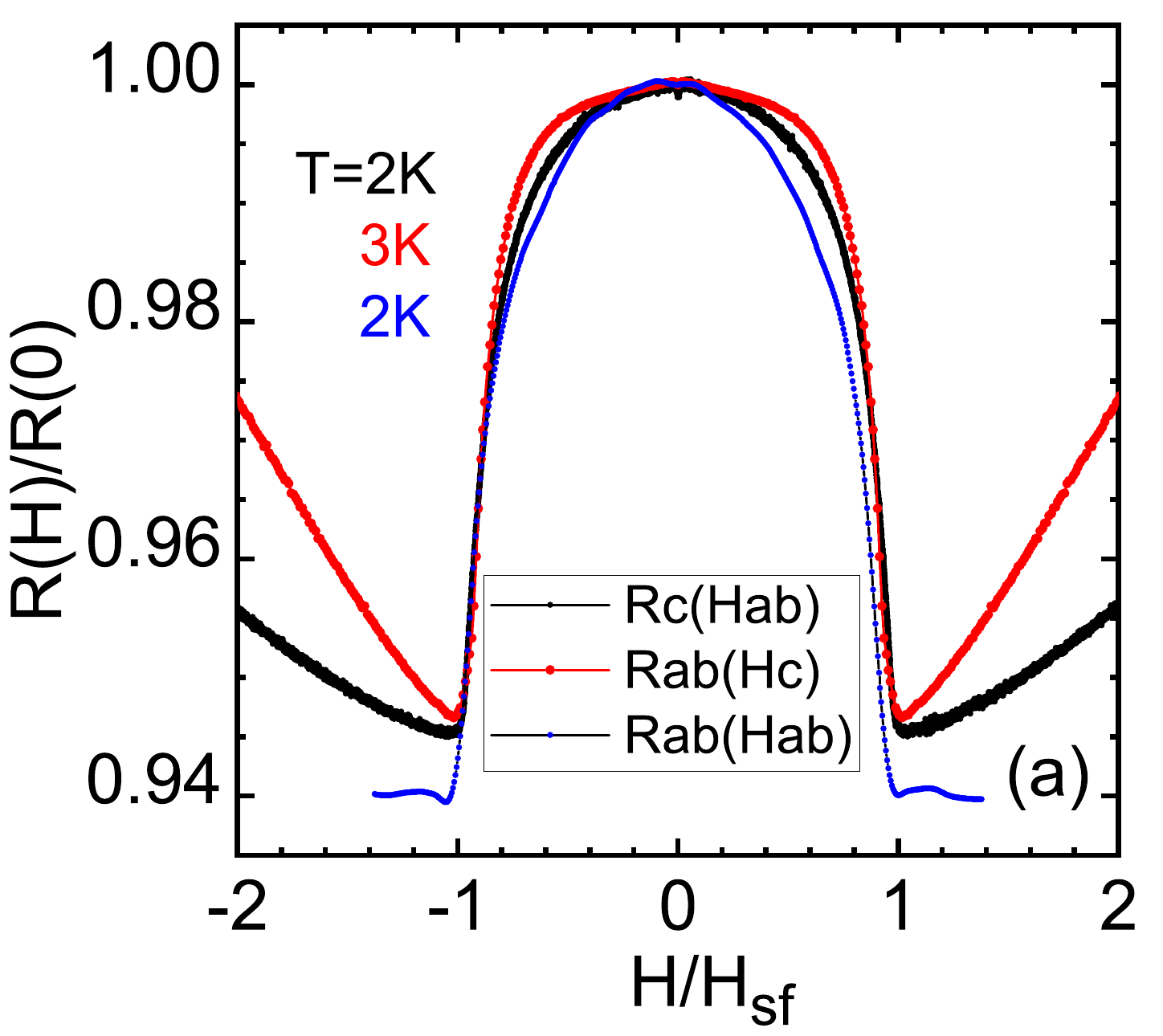}
\includegraphics[width=120pt]{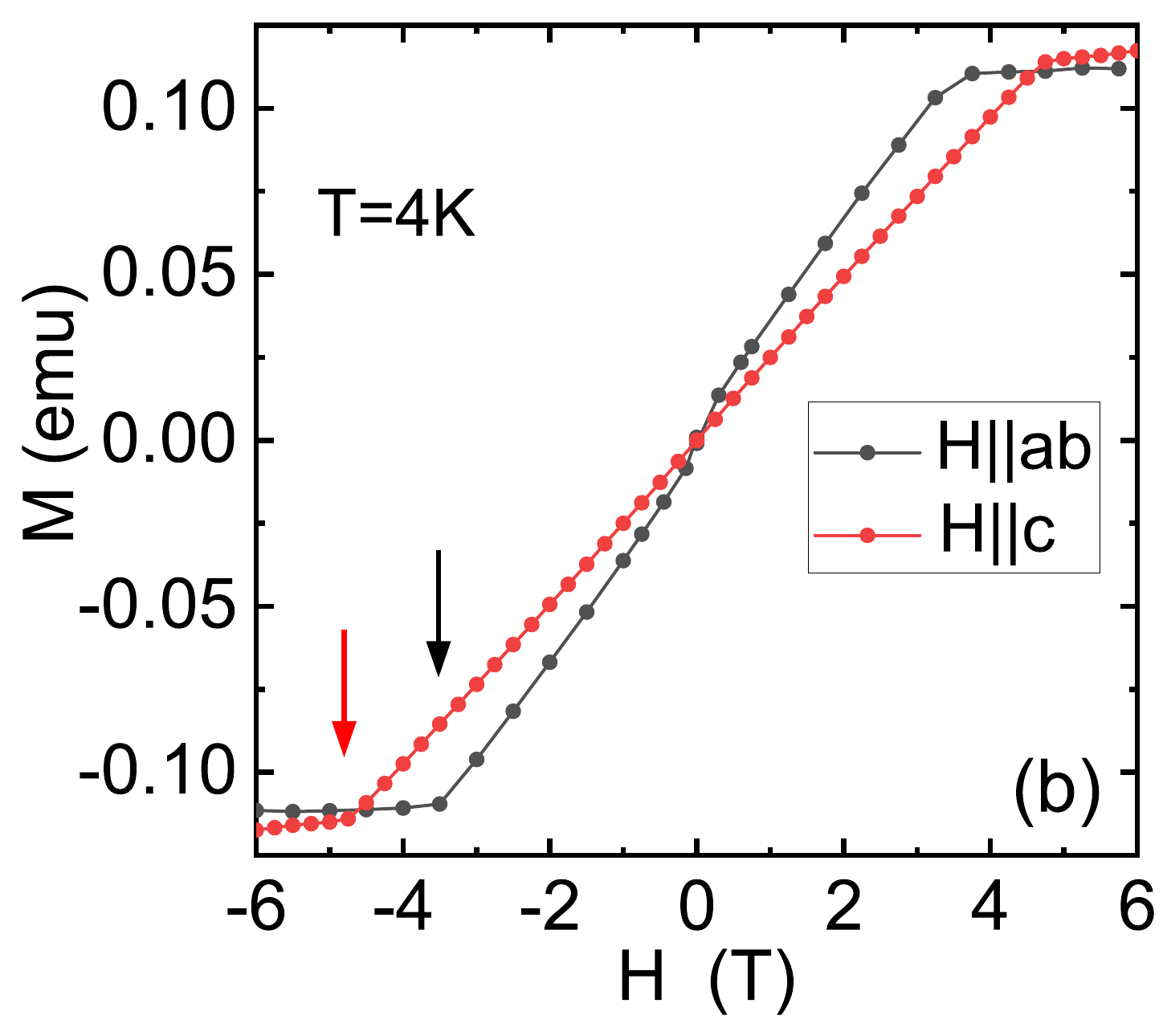}
\caption{(a) Examples of the normalized magnetoresistance  $R(H)/R(0))$ vs normalized field $(H/H_{\rm sf})$ for two orientations of the bias current, in-plane $R_{ab}$ and perpendicular to the plane $R_c$, and for two magnetic field directions, $H\|(ab)$ and $H\|c$.  Data are taken at $T=2-3$K. (b) Magnetization $M(H)$  dependences  for two field orientations, $\mathbf{H}\|ab$ and $\mathbf{H}\|c$. The nonlinearity of $M(H\|ab)$ in low fields is related with   spin canting and spin-flop \cite{PRB_tbp}. Vertical arrows depict the complete spin polarization field $H_{\rm sf}$.
}
\label{fig:parabolic_MR}
\end{figure}

At $H>H_{\mathrm{sf}}$ the obtained correction (\ref{drho}) disappears, and
one returns to the usual positive magnetoresistance in multiband conductors due to
impurity scattering, which is parabolic at low field when $\omega_c \tau \ll 1 $ \cite{PRB_tbp}: 
\begin{equation}
\rho_{zz}^{m}\left( H\right) /\rho_{zz}^{m}\left( 0\right)= 1+\omega_c
^{2}\tau ^{2},\ \ \omega_c \tau \ll 1,  \label{RMzB}
\end{equation}
where $\omega_c =eH/(m^{\ast}c)$ is the cyclotron frequency. Combining Eqs. (\ref{drho}) and (\ref{RMzB}) gives the schematic MR curve 
\begin{equation}
\frac{\rho \left( H\right) }{\rho \left( 0\right) }\approx 1-\frac{\left(
	\Delta E_{ex}/2t_{0}\right) ^{2}}{1+2\left( \Delta E_{ex}/2t_{0}\right) ^{2}} 
\frac{H^{2}}{H_{\mathrm{sf}}^{2}}+\left[ \omega _{c}(H)\tau \right]	^{2},  \label{RMzF}
\end{equation}
illustrated in Fig.~\ref{FigRB}a, which resembles very much the experimental
observations shown in Fig.~\ref{fig:parabolic_MR}. The first-order spin-flip phase 
transition leads to a jump or much faster decrease of AFM order parameter 
$M_{AFM}\left( H\right) $ at $H \approx  H_{\mathrm{sf}}$ than the continuous dependence 
given by Eq. (\ref{M}). Hence, at $H$ close to spin-polarization field $H_{\mathrm{sf}}$
the resistivity drops faster than in Eqs. (\ref{drho}), (\ref{RMzF}),  or in Fig. \ref{FigRB}a. 

In usual 3D AFM metals the proposed NMR mechanism is very weak, because the
ratio $\gamma=\Delta E_{ex}/2t_{0}\ll 1$. Indeed, $\Delta E_{ex}\lesssim
0.1$eV, while $2t_{0}\sim 1$eV is comparable to the bandwidth. However, in
strongly anisotropic layered materials with A-type AFM order the ratio $%
\gamma=\Delta E_{ex}/2t_{0}\sim 1$, because the
interlayer transfer integral $t_{0}\lesssim 0.1$eV in van-der-Waals or other layered compounds is also small.

{\bf Comparison of the model with experimental data}. To compare the presented theory with experimental data,  we performed magnetization and magnetoresistance 	measurements with EuSn$_2$As$_2$ bulk single crystals.

The normalized $R(H)/R(0)$ magnetoresistance and magnetization $M(H)$ dependences 
 are shown in Fig.~\ref{fig:parabolic_MR}  for two field orientations $H\|c$ and $H\|ab$ at low temperature $T\approx 2$\,K,  $T\ll T_N\approx 24$\,K.
The most striking here is the  NMR hump that is the focus of our study.   NMR terminates sharply at the 
full spin polarization field $H_{\rm sf}$, and the hump magnitude is roughly the same for $H\|c$ and $H\|ab$. 

For the majority of samples,  NMR indeed varies about quadratically with field  as Fig.~\ref{fig:parabolic_MR} shows. Whereas the NMR   magnitude is about  the same for all studied  samples, 
in low fields NMR looks somewhat more flattened. Scaling analysis of the ``flattened''  NMR \cite{PRB_tbp} reveals that it may be decomposed into two terms. Besides the main {\em negative} parabolic term that extends to $H_{\rm sf}$, there is a minor {\em positive} term  that starts   parabolically from $H=0$  and than disappears as field increases, already at $H\approx 2 - 3$\,T$< H_{\rm sf}$. Its possible origin  is discussed further. We therefore consider the quadratic-type NMR (see Fig.~\ref{fig:parabolic_MR}) to be generic dependence (NQMR) and compare it with theory.

As field increases and exceeds $H_{\mathrm{\rm sf}}$, the NQMR hump changes sharply to the  conventional smooth  parabolic rise.
We highlight that the sharp transition between the negative (NQMR) and positive (PMR) magnetoresistance (i.e. the $R(H)$ minima) coincides with the sharp magnetization saturation at $H=H_{\mathrm{sf}}$ for both field
directions (Figs. \ref{fig:parabolic_MR}a,b). 
This is consistent also with previous studies \cite{chen_ChPhysLet_2020, li_PRB_2021, sun_SciChin_2021}.
Comparing this data with the presented first-order in $(\omega_c\tau)^2$ model, 
we conclude that the model correctly captures the main features: approximately parabolic NMR, 
its magnitude, and a sharp transition to the conventional PMR. 

 In our measurements   on EuSn$_{2}$As$_{2}$ crystals 
	with A-type AFM order, the magnetoresistance   $\delta \rho /\rho$ drops by about $5-6\%$, as shown 
in Fig. \ref{fig:parabolic_MR}. According to Eqs. (\ref{drho}) and (\ref%
{delta}) for the band-polarized case $\Delta E_{\zeta}>\varepsilon _{F}$ this NQMR drop  corresponds to 
\begin{equation}
\Delta E_{ex}/4t_{0}=\sqrt{\delta \rho /4\rho }\approx 0.11.  \label{ratio}
\end{equation}%
We now compare this ratio with our DFT calculations and ARPES data for $\Delta E_{ex}$ and 
$4t_{0}$. 
From the DFT calculations we found $\Delta E_{ex}\approx 30$\,meV level splitting
for EuSn$_{2}$As$_{2}$ \cite{PRB_tbp}. Substituting this to Eq. (\ref%
{ratio}) one gets $t_{0}\approx 67$meV. The 
ARPES data \cite{li_PRX_2019, PRB_tbp} do not have sufficient energy resolution to measure the bilayer 
splitting and $t_{0}$ directly. However, the $t_{0}$ value can also be 
roughly estimated from the observed resistivity anisotropy $\rho _{zz}/\rho
_{xx}\approx 100$ \cite{PRB_tbp} and the width of the in-plane energy band $4t_{x}\approx 1.9$%
eV, which is taken from our ARPES data and from the DFT calculation. As a
result, we obtain the lowest estimate $t_{0}\approx t_{x}\sqrt{\rho
_{xx}/\rho _{zz}}\approx 50$meV, which is in a reasonable agreement with $%
t_{0}\approx 67$meV determined from Eq. (\ref{ratio}).

Now we check whether our assumption of complete polarization is valid for
EuSn$_{2}$As$_{2}$. The DFT calculations and ARPES data for EuSn$_{2}$As$%
_{2}$ give rather small Fermi energies $\varepsilon _{F}$ for all
Fermi-surface pockets \cite{PRB_tbp}. According to the DFT calculations, there are two
electron bands with $\varepsilon _{F}\approx 55$ meV and $\varepsilon
_{F}\approx 100$ meV and two hole bands with $\varepsilon _{F}\approx 65$
meV and $\varepsilon _{F}\approx 80$ meV. In order to test whether our assumption of the
complete band polarization is applicable to EuSn$%
_{2}$As$_{2}$, these Fermi energies must be
compared with the band splitting 
\begin{equation}
\Delta E_{\zeta }=E_{-}-E_{+}=2\sqrt{\Delta E_{ex}^{2}/4+t_{0}^{2}}.
\label{DEpm}
\end{equation}%
Substituting $t_{0}\approx t_{z0\rho }\approx 50$meV and $%
\Delta E_{ex}\approx 30$meV to Eq. (\ref{DEpm}) we get the band splitting $%
\Delta E_{\zeta }\gtrsim 100$meV$>\varepsilon _{F}$ for all Fermi-surface
pockets. 
 We conclude, EuSn$_{2}$As$_{2}$ corresponds to the completely
band-polarized case for all electronic bands at the Fermi levels, and the above 
analysis, including  Eq. (\ref{drho}), is applicable to the  experimental data  for EuSn$_{2}$As$_{2}$.  

\begin{figure}[tbh]
	\includegraphics[width=0.48\textwidth]{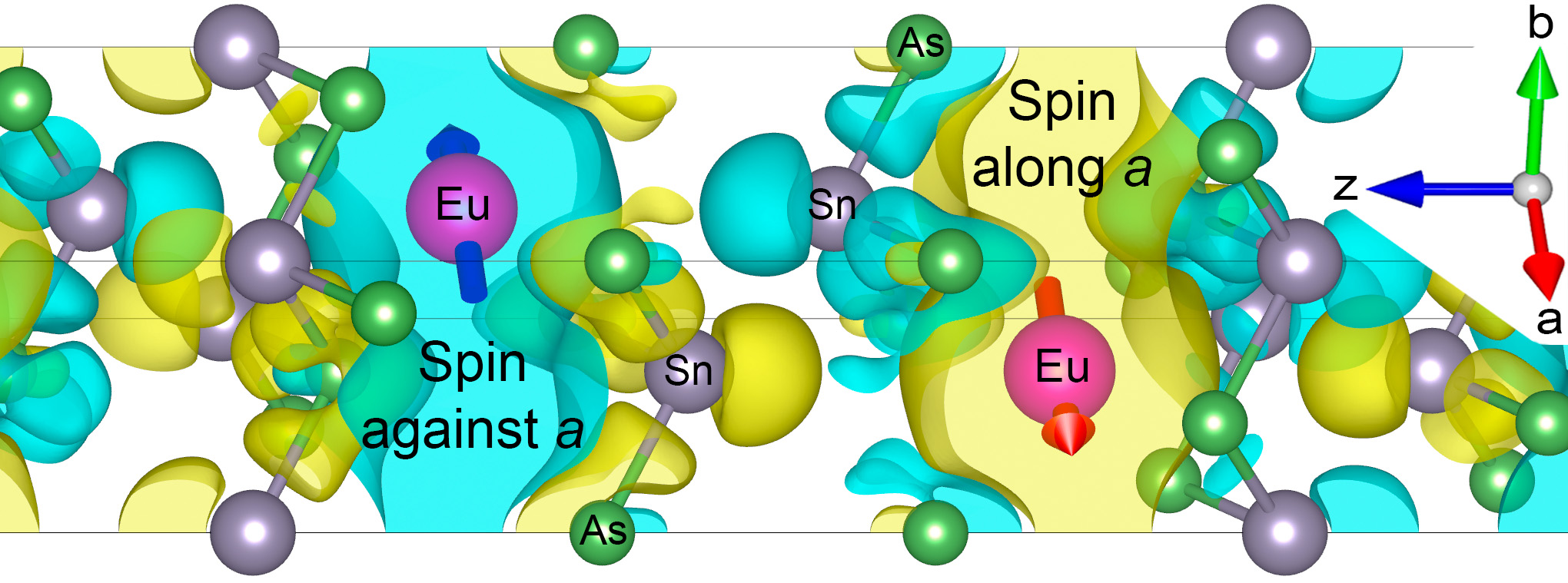}
	\caption{Calculated spin-polarized electron density distribution. 
			The yellow (turquoise) isosurfaces correspond to fixed differences of charge density with spin along (against) the $a$ axis of AFM ordering of Eu spins. The neighboring SnAs layers have opposite spin polarization of electrons at the Fermi level thus substantiating our model.}
	\label{FigDFT}
\end{figure}

To substantiate our model of opposite spin polarization of conducting electrons on neighboring SnAs layers we performed the ab-initio calculation of spin-polarized electron density distribution using the density-functional theory (DFT). Our results are shown in Fig. \ref{FigDFT}, where the yellow (turquoise) isosurfaces limit the real-space areas filled with the same color of positive (negative) electron spin density exceeding a fixed absolute value. As we assumed in our model and illustrated in Fig. \ref{FigWF}, the neighboring SnAs layers indeed have opposite spin polarization of conducting electrons. This spin polarization in EuSn$_2$As$_2$ is small, $\sim 0.1$\% of total electron density, but it is peaked at the Fermi level and strongly affects the electron scattering rate according to the above analysis. 

We highlight  that the proposed NMR mechanism also applies to multiband conductors, where at least one band is 
completely polarized, or even  to those where at the Fermi level the DoS for two subbands 
 splitted by $\Delta E_{\zeta }$ differ considerably. 
However, if all bands are polarized, the described NMR effect becomes stronger and may approach its   theoretical 
 limit of 50\%  at $\gamma\gg 1$, shown in Fig. \ref{FigRB}b, when the resistivity drops by half.

{\bf Discussion}\\
We assess below the potential influence  of other possible mechanisms on  NMR in layered AFM semimetals.
 
 (i) Electron-magnon scattering. 
For antiferromagnets, MR was found to be  {\em positive} 
\cite{yamada-takada_1973-1, yamada-takada_1973-2, note-yamada}, 
 and for $T\ll T_N$ to increase quadratically with field. 
The magnetoresistance   originates from a field-induced increase in  spin fluctuations.
This theoretical result is confirmed by a large  number of experimental studies, e.g. on antiferromagnetic Mn$_2$Au \cite{bodnar_PRA_2020} and PrB$_6$ \cite{ali_JAP_1987}.

 Probably, this mechanism is responsible for a minor flattening of  NQMR in low fields, that is observed for 
 the studied EuSn$_2$As$_2$ samples (see Fig.~\ref{fig:parabolic_MR}a). 
 Indeed, adapting  the results of Refs.~\cite{yamada-takada_1973-1, yamada-takada_1973-2} to our case of the easy-plane antiferromagnet  \cite{note-yamada} we anticipate:
 $ \Delta\rho(H)/\rho(0) \propto (\mu_BH_{c1}/T)^2  \left(H/H_{c1}\right)^2 $ for $\mathbf{H} \|ab$, and $\Delta\rho(H)/\rho(0) \propto (T/T_N)^2  \left(H/H_{c2}\right)^2$ for $\mathbf{H}\|c$.
 Here $H_{c1}$ is the spin-flop field, and $H_{c2}=H_{\rm sf}\sim 5$\,T. Since $H_{\rm sf}$ is much larger than $H_{\rm c1}$, it  follows, that for the $\mathbf{H}\|c$ orientation PMR should extend to much higher fields $\sim H_{\rm sf}$ and to raise with temperature.  For the $\mathbf{H}\|ab$ orientation, PMR is pronounced only in low fields $\sim H_{\rm c1}$ and  decays with increasing $T$.
 This behavior is qualitatively consistent with flattenning of the measured $R(H)$ dependences shown in Fig.~\ref{fig:parabolic_MR} for various field orientations. The magnon scattering might also  contribute to the transformation  with temperature raising  of NQMR to a positive magnetoresistance (for more detail, see \cite{PRB_tbp}).

In contrast, for ferromagnets the magnetoresistance related with  magnon
scattering is negative and originates from the magnetic-field-induced suppression of scattering.
This type of  low temperature NMR was reported for 
ferromagnetic DyNiBC, GdNiBC, HoNiBC, and TbNiBC \cite{fontes_PRB_1999}.
We note that the sudied in our work EuSn$_2$As$_2$ compound is stoichiometric,  antiferromagnetic in the low temperature range of interest (on a large mesoscopic scale $> \lambda_F$), and  has no FM impurities.

(ii) Kondo-scattering, i.e. 
the scattering of conduction electrons  due to randomly located  magnetic impurities. 
This scattering results in a characteristic minimum in electrical resistivity with temperature.
In our case, firstly, no such a minimum or even a tendency is observed \cite{PRB_tbp}, 
and secondly, there are no magnatic impurities in the stoichiometric EuSn$_2$As$_2$ compound grown from high purity raw materials.
   
(iii) 
In ferromagnetic oxides, such as  Ln$_{1-x}$Ca$_x$MnO$_3$, the magnetoresistance is negative  and 
also correlates with magnetization \cite{hundley_APL_1995}:
the magnetoresistance in this case is
phenomenologically described as $R(H,T)\propto \exp[-M(H,T)/M_0]$ 
and  is explained by polaronic hopping  transport
below the ferromagnetic ordering temperature $T_c$. 
The model of polaron hopping transport in insulating oxides is evidently inapplicable to the semimetallic and well conducting EuSn$_2$As$_2$ characterized by the diffusive- rather than hopping-type transport.

(iv) Giant magnetoresistance (GMR).	The observed NQMR, at first glance, is reminescent of 
GMR in superlattices of Fe/Cr where resistance decreases as magnetization in the neigbouring Fe-layers turns with external field from antiparallel to parallel \cite{baibich_PRL(1988)}.
However, GMR is known to develop for charge transport across the layers, while NQMR under study is fully isotropic.  

(v) The chiral anomaly and nontrivial topology are often considered as the origin of the NQMR, 
e.g. in EuIn$_2$As$_2$ and MnBi$_2$Te$_4$. Regarding the investigated compound  EuSn$_2$As$_2$, our  
ARPES measurements \cite{PRB_tbp} didn't reveal band crossings and Dirac points in the close vicinity of the Fermi level.
The magnetotransport measurements also showed that  $\delta R(H)/R(H)$ is almost independent (within 10\%) of the
field direction $H\|ab$ or $H\|c$, of the charge transport direction,
in-plane $R_{ab}(H)$ or normal to the plane $R_c(H)$ (see Fig.~\ref%
{fig:parabolic_MR} and also \cite{PRB_tbp}).  NQMR is also independent of the angle between the 
electric and magnetic field in the easy $ab$-plane \cite{PRB_tbp}.  These results exclude topological origin of NQMR in EuSn$_2$As$_2$.
The  independence of NQMR on the sample thickness 
(from 60\,nm to 0.2\,mm) \cite{maltsev_tbp} indicates that NQMR does not come from scattering by large-scale
macroscopic lattice defects (such as misfit dislocations, domains,
misoriented grains) and  the sample $ab$-plane bending, since these
imperfections are expected to be different for the bulk crystal and for
exfoliated flakes a few monolayer thick.

 Summarizing the above brief review, we conclude that the previously known and discussed  mechanisms cannot explain the observed NQMR in the layered topologically trivial AFM  semimetals. 
Nevertheless, we believe that the
flattenning of  NMR in low fields may be caused by the positive 
MR contribution from electron-magnon scattering. 

 The approximately quadratic negative magnetoresistance  measured in this work is
similar to that observed earlier in other studies with EuSn$_2$As$_2$ 
\cite{sun_SciChin_2021, li_PRB_2021}, 
and with  sister materials -- EuFe$_2$As$_2$ 
\cite{jiang_NJP_2009}, and EuSn$_2$P$_2$ \cite{gui_ACSCentSci_2019}.  
We, therefore, believe that our model is applicable to the listed materials.
 The similarity in NMR also  indicates insignificance 
of such features as the existence of Dirac states close to $E_F$ in  EuSn$_2$P$_2$  \cite{gui_ACSCentSci_2019},   
a stripe SDW in the Fe chains   and a more complex biquadratic coupling between  Eu- and Fe-atoms in EuFe$_2$As$_2$.

There are other more complex behaviors of magnetoresistance in more complex layered materials, for example, in  
ferromagnetic Weyl semimetal  EuCd$_2$As$_2$ \cite{roychowdhury_AdvSci_2020}, in axion insulator EuIn$_2$As$_2$ 
\cite{yu_PRB_2020, yan_PRR_2022}, and AFM topological insulators EuMg$_2$Bi$_2$ \cite{marshall_JAP_2021},  MnBi$_2$Te$_4$ 
 and MnBi$_4$Te$_7$ \cite{ge_NatSciRev_2020, tan_PRL_2020}. Obviously, these cases are beyond the framework 
 of the considered model.

The proposed model helps to extract useful information about the electronic structure of AFM compounds from measurements  of the NQMR effect. Indeed, according to Eqs. (\ref{drho}) or (\ref{RMzF}), the NQMR magnitude gives the parameter $\gamma =\Delta E_{ex}/2t_{0}$, i.e. the ratio of exchange energy splitting to the hopping amplitude between AFM sublattices. 
The sharp NQMR-minimum also helps determine the magnetic field $H_{\rm sf}$ of complete spin polarization 
without magnetization measurements.

{\bf To conclude,} we clarified the origin of the \emph{negative parabolic}
magnetoresistance observed in layered AFM semimetals. Specifically,  we 
suggested a novel type of  magnetoresistance mechanism and
developed a theory describing parabolic NMR over a wide field
range up to complete spin polarization. In the
proposed theory, the negative magnetoresistance originates from the 
violation of $\hat{Z}_2$ symmetry and the corresponding stronger localization of 
electron wave functions on one of the two AFM sublattices depending on electron spin.
In the analyzed semimetals these AFM sublattices are 
 oppositely polarized layers of magnetic ions in the host lattice.
The proposed mechanism of the parabolic NMR  is generic 
for the layered 
AFM van-der-Waals semimetals;  its strength increases 
as the ratio of the exchange splitting to the transfer integral increases. 

The proposed model agrees qualitatively with the available data for layered AFM semimetals, such as 
EuSn$_2$As$_2$, EuFe$_2$As$_2$, EuSn$_2$P$_2$. This agreement is a strong evidence that NMR in 
the layered AFM semimetals is an intrinsic property, irrelevant to defects, domains, and other sample-specific disorder.

\medskip
{\bf Acknowledgements.}
AVS, OAS,  and VMP were supported within  State Assignment of the research at LPI.
PDG acknowledges State Assignment \# FFWR-2024-0015, NUST “MISIS” Grant \# K2-2022-025 and “Basis” 
Foundation for Grant \# 22-1-1-24-1. NSP, IAN and IRS acknowledge  partial support of the State Assignment
\# 124022200005-2 of Institute of Electrophysics and \# 124020600024-5 of Institute of Solid State Chemistry 
UB RAS. AVS, OAS,  VMP, NSP and IAN  acknowledge support 
from RSCF (grant \# 23-12-00307). LV acknowledges support from the Leibniz Gemeinschaft through the Leibniz Competition.
Experimental work was partly done using equipment of LPI Shared facility Center.

\end{document}